\documentclass[12pt,preprint]{aastex}

\usepackage{natbib}
\slugcomment{}

\shorttitle{Sub-surface Flow in AR 11158}
\shortauthors{Jain et al.}

\begin{document}


\title{Divergent Horizontal Sub-surface Flows \\ within Active Region 11158}

\author{Kiran Jain, S.C. Tripathy and F. Hill}
\affil{National Solar Observatory, 950 N Cherry Av., Tucson, AZ 85719, USA}
\email{kjain@nso.edu, stripathy@nso.edu, fhill@nso.edu}



\begin{abstract}

We measure the horizontal subsurface flow in a  fast emerging active region (NOAA 11158)
 using the ring-diagram technique and the HMI
 high-spatial resolution Dopplergrams. This active region had a complex magnetic structure 
 and displayed significant changes in the
morphology during its disk passage. Over the period of six days from 2011 February 
11  to 16,  the temporal variation  in the magnitude of total velocity is found to
follow the trend of  magnetic field strength. 
We further analyze regions of individual magnetic  polarity within AR 11158  and 
 find that the horizontal velocity components in
these sub-regions have significant variation with time and depth.
The leading and trailing polarity regions move faster than the
mixed-polarity region. 
Further, both zonal and meridional components have opposite signs for
trailing and leading polarity regions at all depths showing divergent flows 
within the active region.  We also find a sharp decrease in the magnitude of 
total horizontal velocity  in deeper layer around major flares.
It is suggested that the  re-organization of magnetic fields
during flares combined with the sunspot rotation decreases the magnitude of 
horizontal flows or  that the flow kinetic energy has been converted
into the energy released by flares. After the decline in flare
activity and  the sunspot rotation, the flows tend to follow the pattern of the magnetic activity.
We also observe less variation in the velocity components near the surface but these tend to 
increase with depth, further demonstrating that the deeper layers are more affected by the topology
of active regions.

\end{abstract}


\section{Introduction}

The launch of {\it Solar Dynamics Observatory} \citep[SDO;][]{sdo}
by NASA in February 2010 initiated new interest in studying the dynamic nature of the 
Sun and its consequences.
The unprecedented high-spatial and high-temporal resolution 
simultaneous/ continuous observations, for the first time,  at several wavelengths ranging from 
photosphere to corona provided new details of various layers in the solar atmosphere 
as well as in the solar interior for both quiet and active Sun.
As this was the time when the Sun was recovering from a relatively extended phase of the solar minimum,
changes in the activity level were being monitored very closely by several instruments in space and 
on the ground. The solar activity level is generally measured by the number of sunspots and/or active regions (ARs)
visible on the solar disk. These are the regions of high magnetic field and also the 
main source regions of eruptive events like flares and coronal mass ejections (CMEs), hence they
serve as best examples for understanding the dynamics of the Sun.

The X2.2 flare on 2011 February 15 was the first flare of its class observed by the SDO. It 
originated from a strong bi-polar active region, NOAA 11158, that emerged on the Earth-facing 
side of the Sun's surface on 2011 February 11 and continued to evolve for several days 
until it produced an X-class flare. This region was also the source of tens of C-class and
several M-class flares, and many CMEs were observed during its disk passage. The evolution
of AR 11158 as measured by number of sunspots, sunspot area and associated flares is given 
in Table~\ref{table1}. In addition to its eruptive nature, the active region had several 
rotating sunspots, i.e. rotating around their umbral center or another sunspot(s) 
within the same active region \citep[e.g.][]{brown03}. Now, with improved observational capabilities, 
many more sunspots of these kind are being identified 
 and used to understand the physical reasons for such rotation. Although the cause 
and effect of rotating sunspots are not yet fully understood, it 
has been argued that the twist and shear in the magnetic field may be responsible for this phenomenon
and may lead to the buildup of energy, which might be later released by the flare \citep{stenflo69}.
It is also suggested that the magnetic twist may result from the disturbance in large-scale 
flows in the solar convection zone and the photosphere or sub-photospheric layers \citep{lopez03}.

While most studies on AR 11158 are confined to exploring conditions in the photosphere and higher
atmosphere \citep[e.g.][]{schrijver2011, petrie2013, wang2014, malanushenko2014, jiezhao2014,Sorriso2015}, 
a very few are available for those below the surface \citep[e.g.][]{gao2012, gao2014}. The former investigations
cover a large variety of studies including  sudden/rapid photospheric motion within the active region, 
magnetic topology, evolution of helicity by shearing motion of flux tubes, and reconstruction of 
various events associated with this region. However, the conditions below the surface were
not examined in detail. \citet{gao2012} investigated the correlation between photospheric
current helicity and subsurface kinetic helicity by analyzing vector magnetograms and subsurface
velocities. The velocities were computed using the time-distance pipeline of HMI
 and analysis
was carried out only for a shallow depth, i.e. 0 -- 1 Mm. In this paper, we investigate the
changes in deeper layers (down to 12 Mm) by exploring velocity fields and their temporal evolution.
The analysis is carried out for the active region as whole and also 
for the leading, trailing and mixed-polarity flux regions separately.

The paper is organized as follows; in Section 2, we describe the data and technique used in
the analysis and the data reduction. The results are presented in Section 3 followed by the
discussion in Section 4. Finally, we summarize our findings in Section 5.

\section{Analysis}
\subsection{Data and Technique}

High-spatial resolution Doppler images from the Helioseismic and
Magnetic Imager \citep[HMI;][]{hmi}  
on board SDO were used in this study. The Dopplergrams were constructed using 
measurements taken in the 6173.3 \AA ~absorption line in the photosphere with pixel size of 0.5 arcsec 
and cadence of 45 s. We applied one of the techniques of local helioseismology, known as ring-diagram analysis,
to calculate flows in the thin layer below
the solar surface. In this technique, high-degree acoustic modes are used to infer the characteristics 
of the propagating waves in localized areas \citep{hill88}. This method has 
been  extensively used to study long-term variations in sub-surface flows \citep{komm13, komm14} as well as
short-term variations in subsurface properties of both active and quiet regions  \citep{baldner11,basu04,
rick08, jain08, jain12, komm11, rajaguru01, mcrs13, sushant08, tripathy13a}. 

In this study, we track 
the regions of interest for 1440 min  using the surface rotation rate \citep{snod84} and apodize each 
tracked area with a circular function. Then a three-dimensional FFT is applied in 
both spatial and temporal directions to obtain a three-dimensional power spectrum that is fitted  
using a Lorentzian profile model \citep{haber00},
\begin{eqnarray}
P (k_x, k_y, \omega) & = & {A \over (\omega - \omega_0 + k_xU_x+k_yU_y)^2+\Gamma^2}  + {b \over k^3} 
\end{eqnarray}
where $P$ is the oscillation power for a wave with a temporal frequency ($\omega$) and
the total wave number $k^2=k_x^2+k_y^2$. There are six parameters to be fitted:
two Doppler shifts ($k_xU_x$ and $k_yU_y$) for waves propagating in the orthogonal 
zonal and meridional directions, the background power ($b$), the mode
central frequency ($\omega_0$), the mode width ($\Gamma$), and the amplitude ($A$).
Finally, the fitted velocities ($U_x$ and $U_y$) are inverted using a regularized least square 
(RLS) method to estimate depth dependence of various components of the horizontal velocity ($V_x$ and $V_y$). 

\subsection{Data Reduction and Systematics}
\begin{figure}   
   \centerline{
\includegraphics[scale=.60]{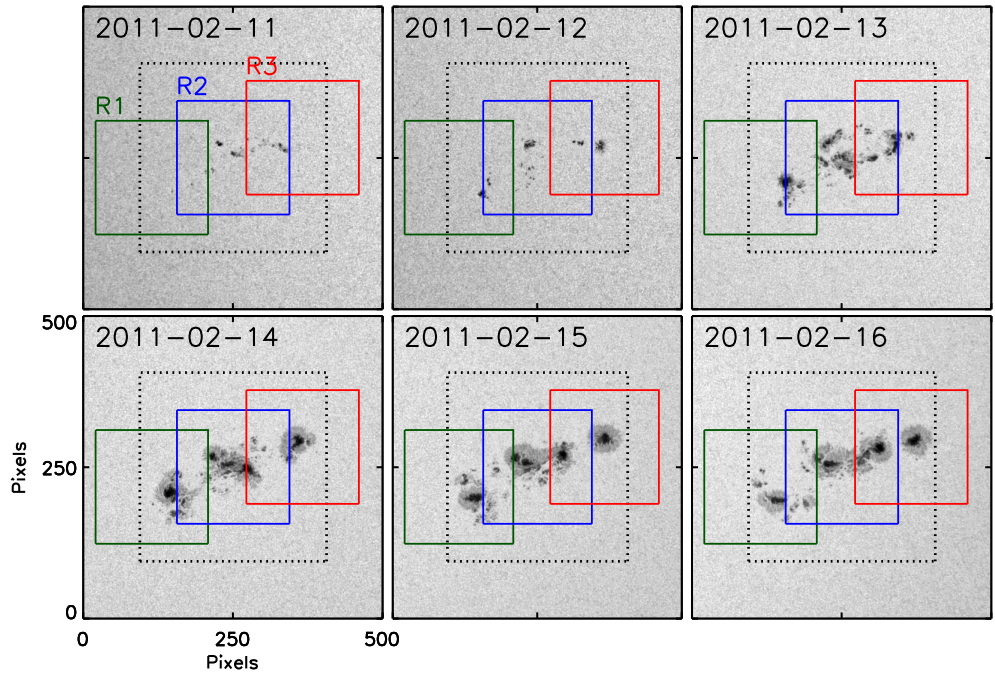}
              }
             \caption{HMI continuum images showing the evolution of AR 11158 for six consecutive days.
             The boxes show various regions covered in this analysis. The 
             12$^{\mathrm{o}}\times$12$^{\mathrm{o}}$ box is shown with black dotted line covering
             the entire active region. Smaller  7.5$^{\mathrm{o}}\times$7.5$^{\mathrm{o}}$ 
             boxes marked with R1 (green), R2 (blue) and R3 (red) cover trailing, mixed
             and leading polarity regions of AR 11158.
    }
   \label{figint}
   \end{figure}

\begin{figure}   
   \centerline{
\includegraphics[scale=.60]{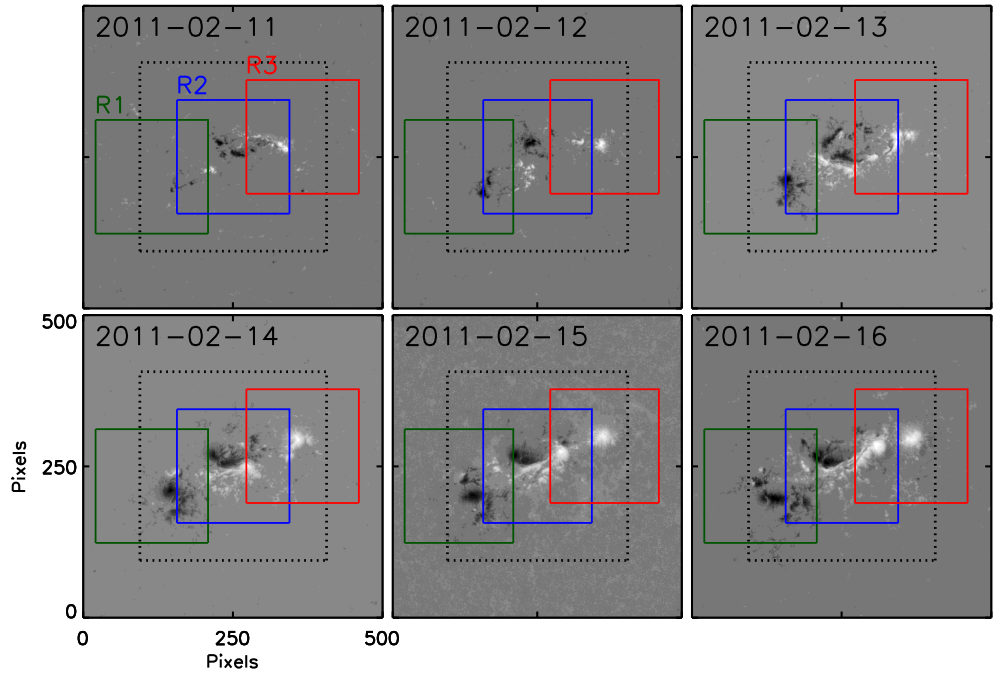}
              }
             \caption{Same as Figure~\ref{figint} but for HMI magnetograms.
    }
   \label{figmag}
   \end{figure}

We analyze Dopplergrams using two sets of regions; the first set comprises
regions of $\approx$12$^{\mathrm{o}}\times$12$^{\mathrm{o}}$  while the second set contains regions
of $\approx$7.5$^{\mathrm{o}}$$\times$7.5$^{\mathrm{o}}$.  These regions are chosen in such a
way that the entire active region is covered in set 1 and three smaller regions are 
considered in set 2 which are dominated by
 (i) trailing magnetic polarity - R1, (ii) mixed polarity - R2, and (iii) leading polarity - R3.
 The evolution of AR 11158 in six consecutive days is shown in continuum intensity
and magnetograms in 
Figures~\ref{figint} and  \ref{figmag}, respectively.
The areas covered by 7.5$^{\mathrm{o}}$$\times$7.5$^{\mathrm{o}}$ regions in these figures
are marked with solid lines while 12$^{\mathrm{o}}\times$12$^{\mathrm{o}}$ regions are shown by dotted lines. 
Details of each region are given in Table~\ref{table2} where locations are tabulated for reference images 
(i.e. the image at the center of each time series). It should be noted that the
Carrington longitude and latitude of each region remain unchanged  which ensures that the areas
studied here do not change with time or the rotation of the Sun. 

In order to investigate 
the temporal evolution of velocity fields  in various time samples, it is necessary to
address the influence of systematics. It has been shown in previous studies that the helioseismic
measurements suffer from a variety of systematic errors.  One of the potential source of errors is 
the center to limb effect \citep[for example,][]{jain13a}.  This effect becomes more prominent 
when we compare properties of various regions at different disk positions as it increases with 
the angle from the disk center. Another systematic effect is the annual variation in the solar 
inclination towards Earth, known as $B_0$ angle \citep[for example,][]{zaatri2006}. In both 
cases, we lose the spatial resolution and  increase the errors in measurements. Thus, these 
systematics need to be corrected in order to infer the true properties of sub-surface layers, 
e.g., the evolution of flows in active regions. Several theoretical as well as empirical 
explanations have been suggested to remove these effects from flows  \citep[and 
references therein]{komm2015}. 

In a different approach to reduce these effects and infer residuals in active regions, we
use quiet regions from the neighboring Carrington rotations and subtract their values from those 
 active regions \citep{jain12}. In this study, we select quiet days from the previous 
Carrington rotation at same heliographic locations as in the case of 
active region. The quiet days corresponding to each active region  are also listed in Table~\ref{table2}.
This analysis is further constrained by the selection of
tile sizes. While selection of 
large tiles leads to a larger population of fitted mode available for inversion that penetrate the deeper layers, it
dilutes the effect of strong magnetic field due to the inclusion of neighboring quiet regions.
In this paper, we present analysis for two  depth ranges; down to  12 Mm for the 12$^{\mathrm{o}}$
tiles and down to 7  Mm for 7.5$^{\mathrm{o}}$ tiles.

\section{Results}
\subsection{Evolution of magnetic field in analysis windows}

\begin{figure}   
   \centerline{
\includegraphics[scale=.80]{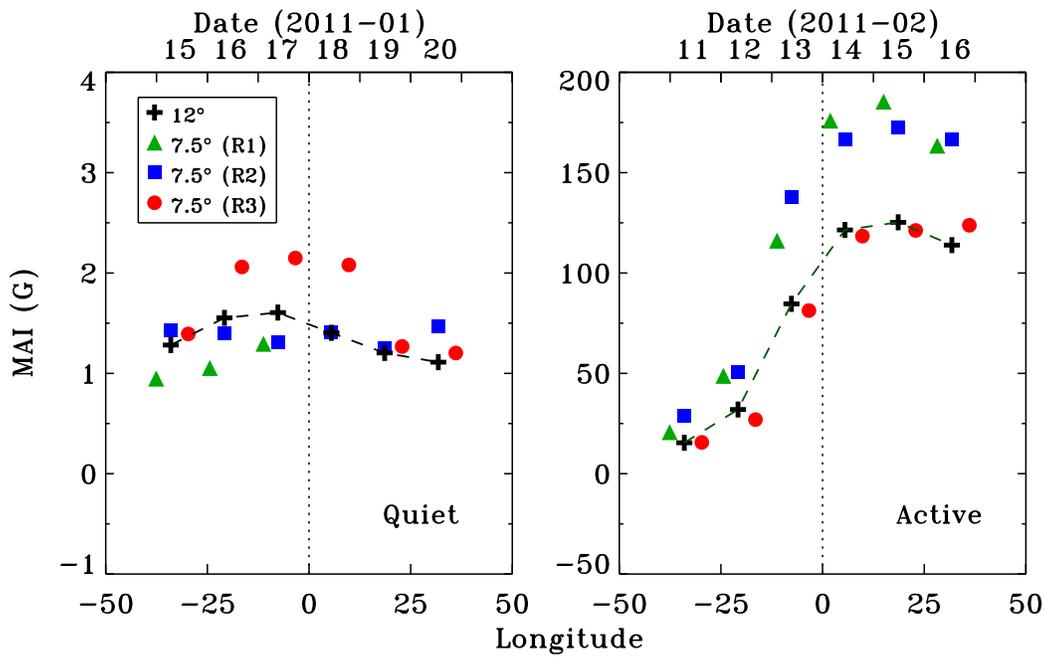}
              }
            \caption{Temporal variation of magnetic activity index (MAI); (left) in quiet regions, and (right) in active region. Values for 
             12$^{\mathrm{o}}\times$12$^{\mathrm{o}}$  regions are joined by the dashed line while those
             for 7.5$^{\mathrm{o}}\times$7.5$^{\mathrm{o}}$ regions are shown by different symbols.}
   \label{figmai}
   \end{figure}

 AR 11158 had a complex magnetic field structure with four major concentrations. 
 The magnetic flux distribution was tilted with
respect to the equator. The leading polarity flux, which was positive, was the most
equatorward and the trailing polarity flux, which was negative, was the most poleward. The
distribution of flux was well aligned with the  Hale-Nicholson law. In order to parameterize 
the evolution of AR 11158 in different analysis windows, we plot the magnetic 
active index (MAI) in Figure~\ref{figmai}, and provide the mean MAI and standard deviation in Table~\ref{table3}.
Values for corresponding quiet regions from the previous Carrington rotation are also included here. 
 The MAI is a proxy for magnetic activity that is calculated by converting line-of-sight magnetogram data into 
 the absolute value \citep{basu04}.  The averages 
 are computed  using the HMI magnetograms  over  the same length of time as in the 
Dopplergram tiles. It is noted that  the AR 11158 was a fast emerging active region and
its morphology changed significantly during the course of its disk passage.  
In contrast to AR 11158, there is  less variation in MAIs in quiet areas with time, hence the 
standard deviations are smaller.   

\subsection{Sub-surface flows in selected quiet regions}
\begin{figure}   
   \centerline{
\includegraphics[scale=.80]{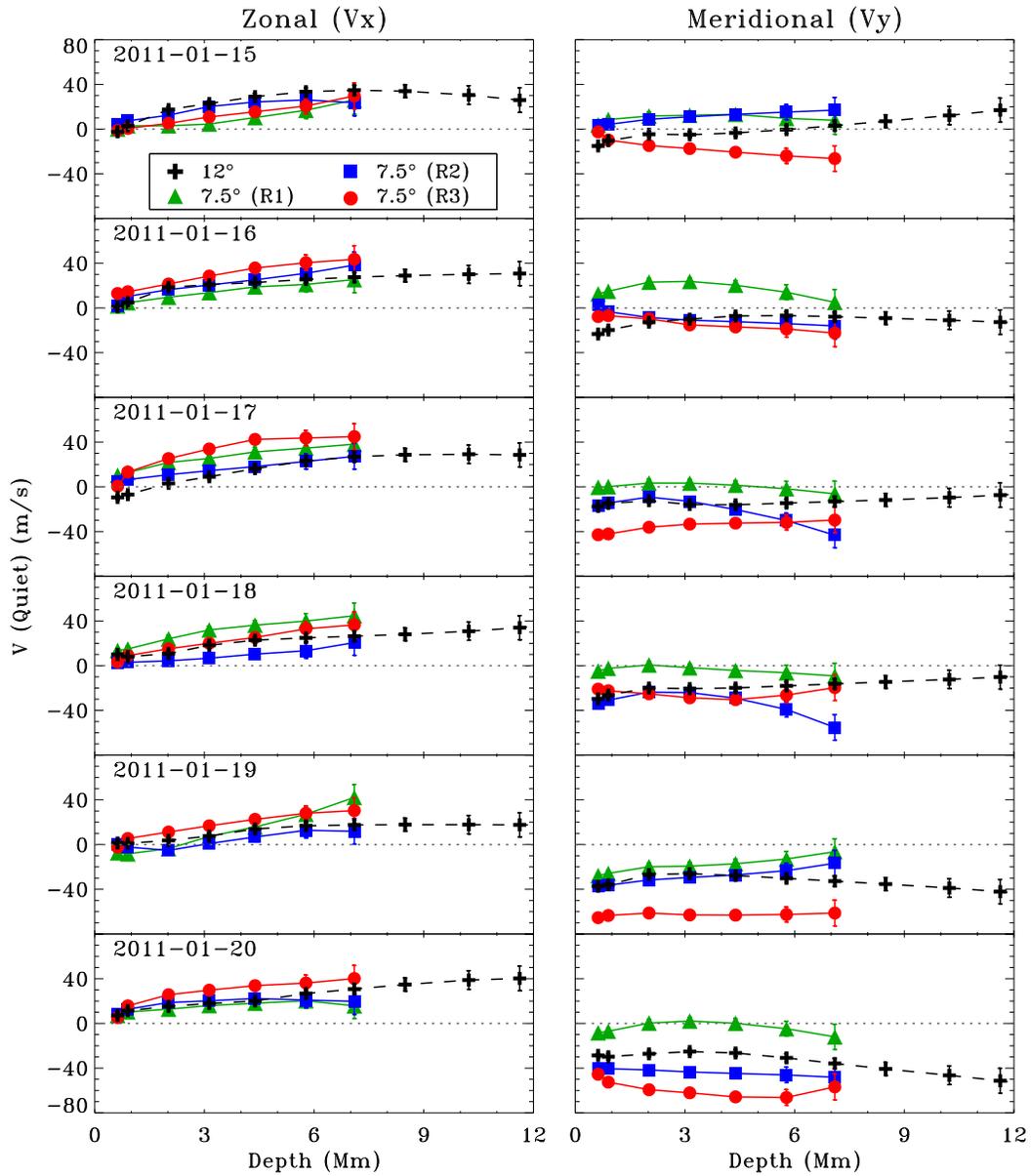}
              }
            \caption{Depth variation of zonal (left) and meridional (right)
            components of the horizontal velocity in quiet regions for six consecutive days. }
   \label{figuq}
   \end{figure}

\begin{figure}   
   \centerline{
\includegraphics[scale=.80]{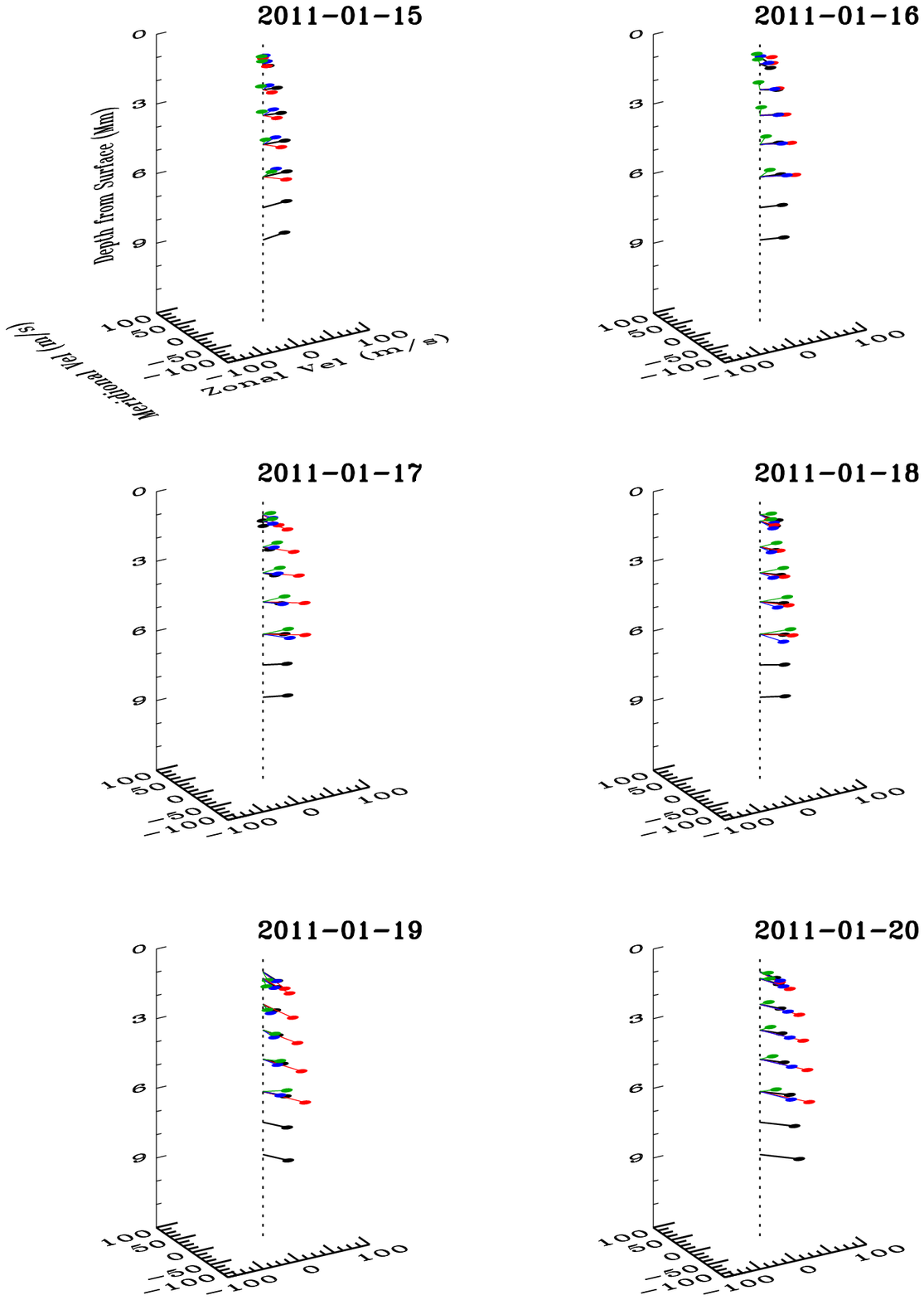}
              }
            \caption{Depth variation of total horizontal velocity in quiet regions. 
             Flows in 12$^{\mathrm{o}}\times$12$^{\mathrm{o}}$ tiles are shown by 
            black while those in regions R1, R2 and R3 are shown by green, blue and red.}
   \label{figboxq}
   \end{figure}

We display zonal and meridional  components  
of the horizontal velocity, $V_x$ and $V_y$,  in different quiet regions as a function of depth in  Figure~\ref{figuq}.
As the presence of strong magnetic field modifies the flow of plasma in the solar interior,
the small variation of the MAI values in quiet regions, as illustrated in Figure~\ref{figmai}, leads to 
the lower temporal variation in flows of these regions. The $V_y$ for different regions are within 1$\sigma$,
however, the magnitude
of $V_x$ is seen to increase with depth which is in agreement with the observations where
rotation rate is found to increase with depth in the outer 5\% of the Sun. The
visible differences in the values of $V_y$ component may be attributed to the different locations of 
regions on the solar surface, hence the center-to-limb effect. However, in most cases, the 
meridional component is negative, i.e. poleward in southern hemisphere. As one may assume similar 
geometric effect on parameters computed at same locations, we use these quiet region values to 
eliminate the systematics and contribution of the background magnetic field (quiet Sun)
from the active region values. The depth dependence of total horizontal velocity $V$ is plotted in
 Figure~ \ref{figboxq} for all six days where $V = (V_x^2 + V_y^2)^{1/2}$. It is clearly visible that
 the variation in $V$ is small with time and also between different regions.

\subsection{Sub-surface flows in AR 11158}

\begin{figure}   
   \centerline{
\includegraphics[scale=.70]{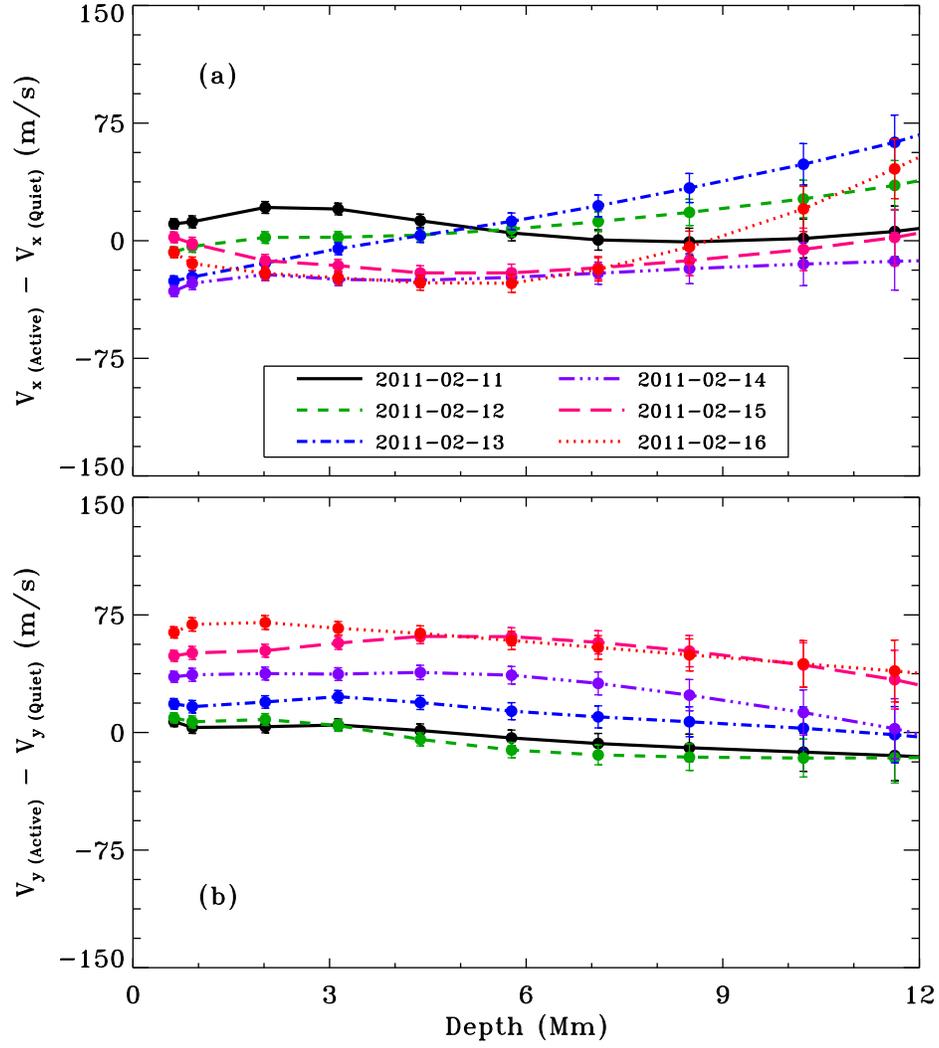}
              }
            \caption{Temporal and depth variation of (a) zonal and (b) meridional components of the
            horizontal velocity in  12$^{\mathrm{o}}\times$12$^{\mathrm{o}}$ tiles
            covering the active region as a whole.}
   \label{fig12a}
   \end{figure}

\begin{figure}   
   \centerline{
\includegraphics[scale=.80]{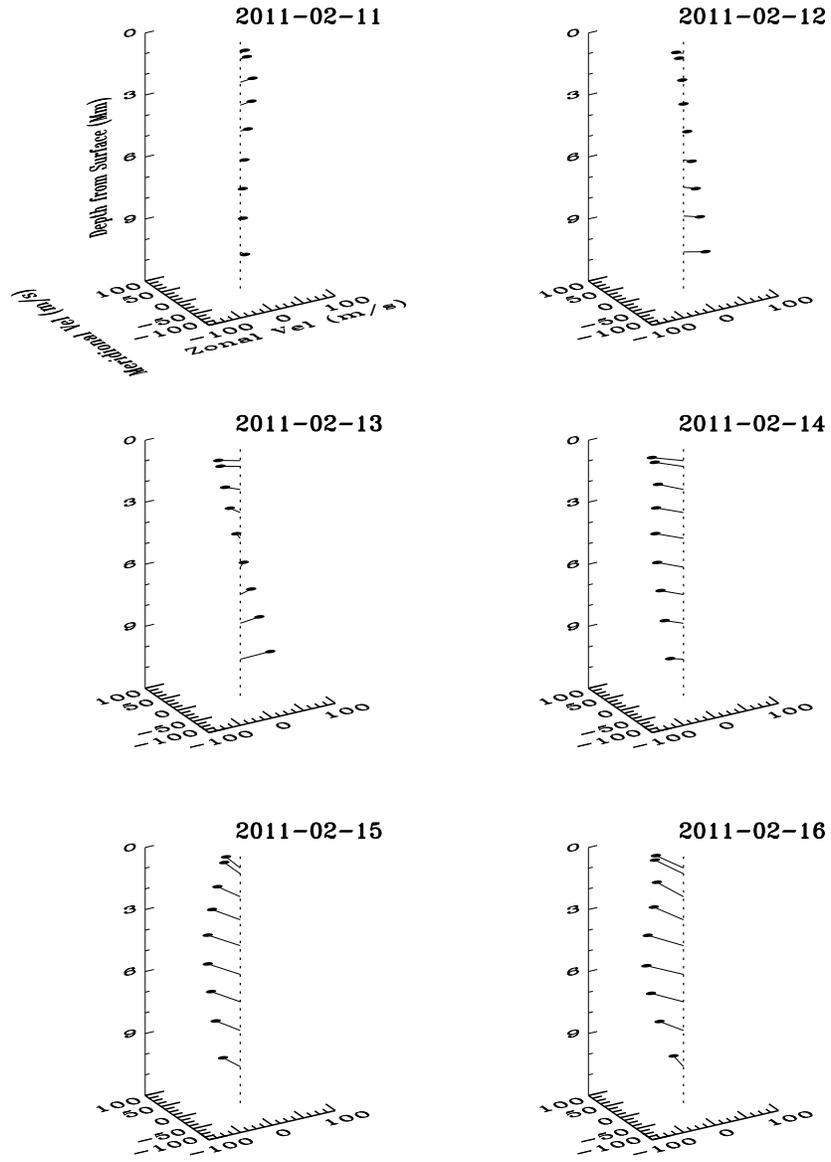}
              }
            \caption{Depth variation of total horizontal velocity vectors in the active region
           for 12$^{\mathrm{o}}\times$12$^{\mathrm{o}}$ 
            tiles. The individual zonal 
            and meridional components are plotted in Figure~\ref{fig12a}.  }
   \label{figbox12}
   \end{figure}

The velocity components, $V_x$  and  $V_y$,   in AR 11158 for 
12$^{\mathrm{o}}\times$12$^{\mathrm{o}}$ regions are  shown in Figure~\ref{fig12a}.  
These are the residual velocities which are obtained by subtracting quiet region
values from those of the active regions. Although the active region went through a rapid growth
during the period of this study,
as shown in Figure~\ref{figmai}, it is interesting to note
that the daily variations in depth profiles were less affected. In six days of the evolution,
the morphology of this active region changed from a simple bi-polar region to a complex region having
intermingled polarities within the single penumbra. There are some changes in the maagnitude of
individual velocity components in deeper layers, however these are within 2--3$\sigma$. 
These results are in agreement with \citet{gao2012} where authors found the magnitudes of
$x$- and $y$-components of the velocity remained stable in upper 1 Mm, and there 
were no coherent and sudden changes. However they did not explore deeper layers where changes
are more visible. 
Since these velocities represent average values for 1440 min, the changes on shorter
time scales can not be described. 
  Figure~\ref{figbox12} illustrates the depth variation of total horizontal velocity
where each panel represents an individual day. Though a small magnitude is visible for the 
first two days at all depths, a change in direction was noticed on the second in shallow layers.
With the increase in magnetic field and other morphological changes within the active region,
a twist in horizontal flow was developed around 6 Mm; the horizontal velocity in shallow layers
was in westward direction and the opposite was seen below this depth down to  12 Mm, the lower-depth limit
in this study. This twist disappears after 
the third day and there was no significant change in velocity for next three days.
 
\begin{figure}   
   \centerline{
\includegraphics[scale=.80]{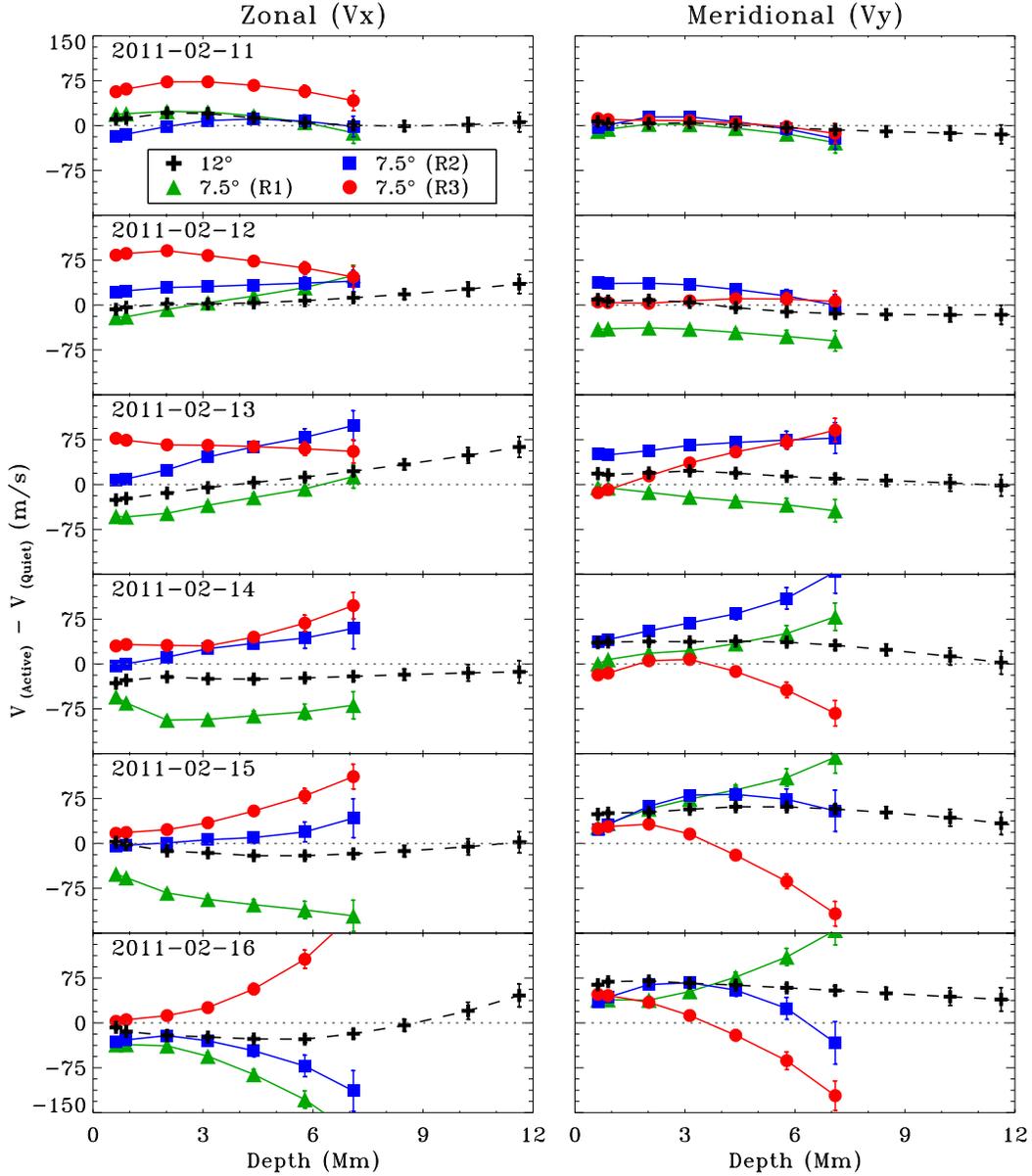}
              }
            \caption{Depth variation of (left) zonal and (right) meridional components of the 
            horizontal velocity in different tiles of the active region for six consecutive days.}
   \label{figua}
   \end{figure}

\begin{figure}   
   \centerline{
\includegraphics[scale=.80]{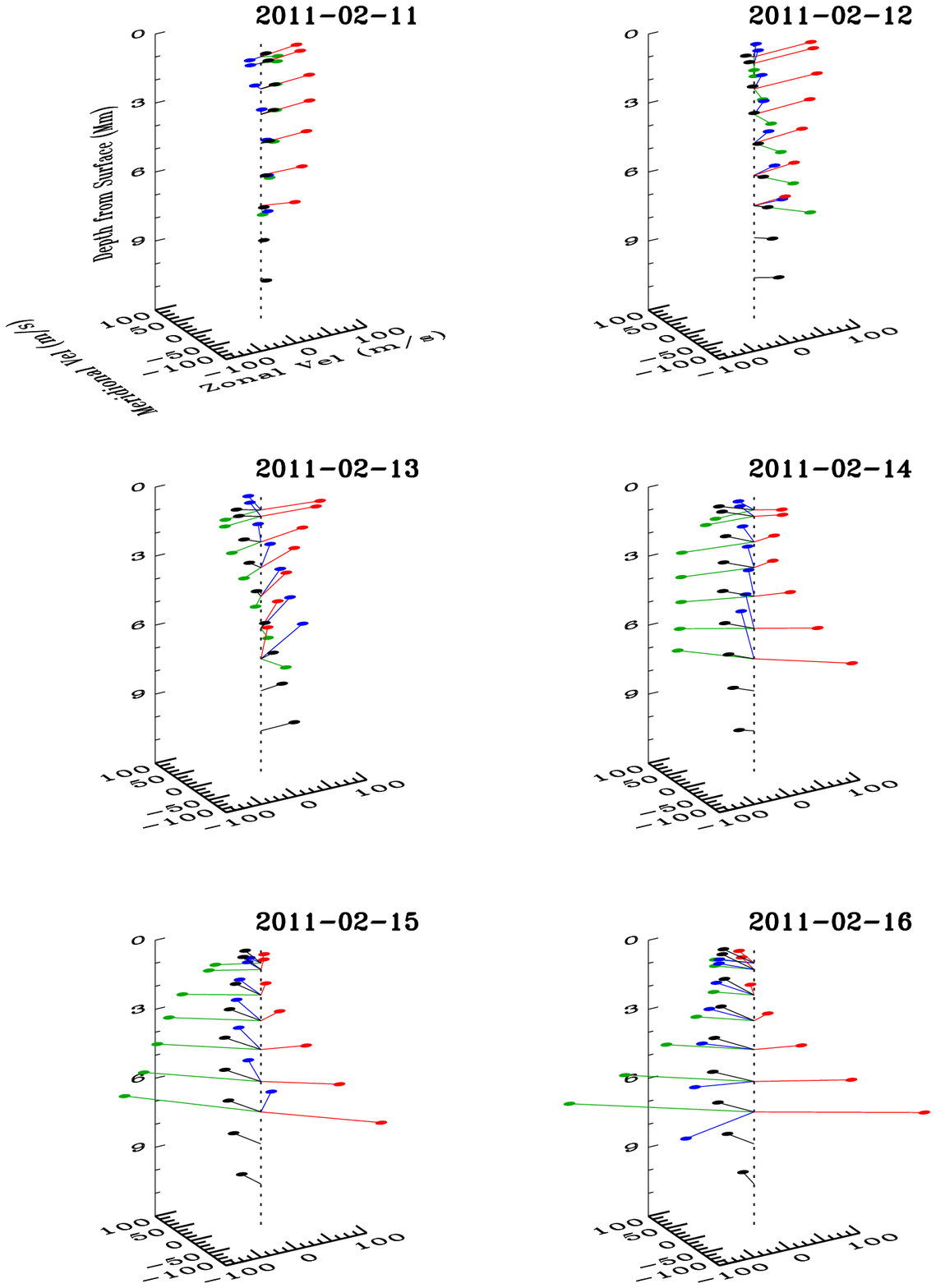}
              }
            \caption{Depth variation of total horizontal velocity in different tiles of the active region.
            Flows in 12$^{\mathrm{o}}\times$12$^{\mathrm{o}}$ regions are shown by 
            black while those in regions R1, R2 and R3 are shown by green, blue and red.}
   \label{figboxa}
   \end{figure}

We further investigate flow patterns within the active region by analyzing
three major sub-regions as described in Section 2.2; the leading polarity, the trailing polarity and
the mixed polarity regions.
The depth variations of the resultant $V_x$  and  $V_y$ in these sub-regions
are displayed in  Figure~\ref{figua}, and the total horizontal velocity in
 Figure~\ref{figboxa}. We also include values for  
12$^{\mathrm{o}}\times$12$^{\mathrm{o}}$ tiles for comparison. 
There are several distinct features in these plots; (i) there is significant variation
in velocity components in 7.5$^{\mathrm{o}}$ tiles as compared to  12$^{\mathrm{o}}$ 
tiles, (ii) the leading and trailing polarity regions move faster than the
mixed-polarity region, (iii) the region-to-region variation is small
when there is no major change in the morphology of active region or sub-regions, 
(iv) both zonal and meridional components have opposite signs for
trailing and leading polarity region at all depths. While $V_x$ and $V_y$ are positive
and negative, respectively, in the leading polarity region, the opposite
is obtained for the trailing polarity region, (v) these velocity components
also change sign in mixed-polarity region after the X2.2 class flare on Feb 15,
and in  the trailing-polarity region after the M6.6 flare on Feb 13 and 
continue to increase their magnitudes until January 16,
(vi) the velocity components have less variation  near the surface but these tend to 
increase with depth demonstrating that the deeper layers are more affected by the topology
of regions. These findings, in context of the possible scenario, are discussed
in the following section.

\section{Discussion}
Several authors have attempted to explain the temporal variation of horizontal flows, the 
origin of sunspot rotation and  their connection with  flares in AR 11158.  \citet{beauregard2012}
measured horizontal motion  with the local correlation tracking (LCT) around the time of 
X2.2 flare on 2001  February 15 using continuum images and magnetograms from
 HMI. They computed horizontal motion averaged over the 1-hour pre- and post-flare 
periods, and found elevated horizontal flow values near the polarity-inversion line 
(located in region R2 of this paper) in the post-flare period.  
\citet{jiang2012} also tracked the evolution of AR 11158 for five days and found that 
the preceeding polarity regions moved faster in westward direction towards 
equator whereas following polarity regions moved slowly in eastward direction. More
recently, \citet{wang2014} investigated the temporal evolution of the photospheric velocity
around  an X2.2 flare and reported a sudden change in shear flows around the flaring
magnetic polarity inversion line. Although the major focus of these studies was on exploring
changes in flow patterns at the major flaring site, they do not provide a complete
picture of the entire active region. Also, all these studies use  the motion of surface features
to determine velocities, hence describe horizontal velocities at or above  the solar surface.

The conditions below the surface can only be measured using the helioseismic techniques.
There have been studies in the past to  investigate flows in the layers very close to the 
surface within the active regions.  It is now accepted that the flows around active regions 
are dominated by inflows into these regions. Using near-surface horizontal velocities 
inferred from the time-distance helioseismology covering a few Carrington rotations in
 Solar Cycle 23, \citet{zhaoetal2004} suggested that the residual rotational velocity of 
 magnetic elements depends on their magnetic field strength: the stronger the magnetic 
 field strength, the faster the magnetic elements rotate relative to the quiet regions. 
 They also found that magnetic elements of the following polarity rotate faster than 
 the leading polarity elements of the same magnetic strength. Later, \citet{svanda2008} 
 confirmed asymmetry between the flows around leading and following polarities of active 
 regions. However, studies using the ring-diagram technique were restricted to analyzing 
 active regions as a whole typically for 15$^{\mathrm{o}}\times$15$^{\mathrm{o}}$ tiles,
mainly  due to the availability of Dopplergrams  with moderate spatial resolution. Using 
this technique, we have analyzed isolated active regions  using tiles as small as 
11$^{\mathrm{o}}\times$11$^{\mathrm{o}}$ to study the temporal variation of sub-surface  horizontal velocity 
in three consecutive Carrington rotations \citep{jain12}. \citet{ hindman2009} also used 
the ring-diagram method to  determine flows in smaller areas of active regions by exploiting 
surface gravity waves ($f$ modes). This approach provided direct measurement of flow velocity 
in the layers where the $f$ mode has significant amplitude, i.e. the layer spanning about 
2 Mm below the surface. All these studies confirmed that the magnetic regions move at 
considerably higher velocities as compared to the quiet regions and the velocities depend 
on the magnetic field strength.  In the present study of AR 11158, we also observe a link 
between magnetic field strength and computed velocities.  However, in addition to the 
magnetic field changes, the active region has an associated characteristic, i.e. the 
sunspot rotation that developed at the  time of increased magnetic field strength. Thus, 
the resultant elevated flow values  are a combination of both effects, i.e., the strong 
magnetic field and sunspot rotation.

\begin{figure}   
   \centerline{
\includegraphics[scale=.70]{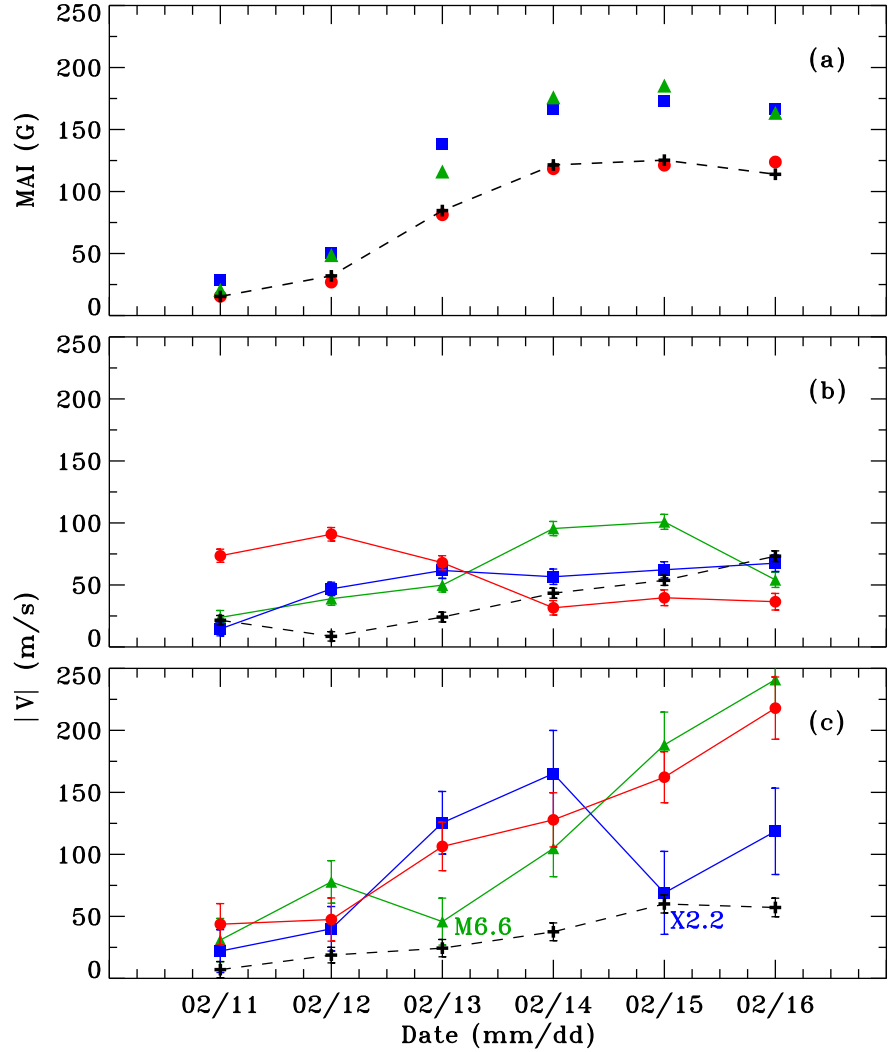}
              }
            \caption{Temporal variation of (a) Magnetic Activity Index (MAI), (b) total horizontal
            velocity at 2 Mm, and (c) total horizontal velocity at 7 Mm in regions R1 (trailing polarity; green triangles),
            R2 (mixed polarity; blue squares) and R3 (leading polarity; red circles).  Major flare events are also marked
            in the bottom panel. For comparison, the variation in 12$^{\mathrm{o}}\times$12$^{\mathrm{o}}$ regions are also shown by 
            black dashed line in each panel.
            }
   \label{figdepth}
   \end{figure}

We further show, in Figure~\ref{figdepth},  examples of the temporal variation of average 
horizontal velocity,  $|{V}|$, in different polarity  regions  at two depths;  one near the 
surface around 2.0 Mm and other at a deeper layer around 7.0 Mm. The variation of MAIs in 
all three regions is also shown in the top panel. Regions R1 (trailing polarity) and R2 
(mixed polarity) have similar variations in MAIs while  region R3 (leading polarity) 
has consistently lower values than the other two regions. As evident from panels (b) and (c), 
the velocity magnitude and errors are larger at 7 Mm as compared to the 2 Mm layer. 
Although there is less variation in the near surface layers (panel b),   we notice 
 two major dips in panel (c); one on February 13 for R1 and another on February 15 for R2. 
 These dips interestingly coincide with two major flares; the M6.6 flare on 
February 13 from R1 and X2.2 flare on February 15 from R2. We assume the  re-organization 
of magnetic fields during flares combined with the sunspot rotation decreased the magnitude 
of horizontal flows at the time of flares. After the decline in flare activity and  the 
sunspot rotation, the flows tend to follow the pattern of the magnetic activity.
This is supported by the study of \citet{sudol05} where they provided strong evidence for
the re-organization of magnetic field during major flares. These findings need to be confirmed 
with a larger set of active regions of similar kind. With
the ongoing operations of SDO/HMI, we anticipate identifying more cases where we can apply the helioseismic
techniques to infer properties/flow fields in the regions of different polarities within active regions.

\section{Summary}

In summary, we measure the horizontal subsurface flow in the  fast emerging AR 11158 
from the HMI high-spatial resolution Dopplergrams using the ring-diagram technique.  
This active region was developed on the Earth-side of solar disk; had a complex magnetic 
structure and displayed significant changes in morphology with time, including several 
rotating sunspots and large flares. Over the period of six days,  the temporal 
variation  in the magnitude of the total velocity is found to be related 
to the changes in magnetic field strength as measured by the magnetic activity index. With 
the rise in magnetic activity, we see an increase in the magnitude of sub-surface flow. 
However, the analysis of individual polarity regions provides several interesting results. 
The horizontal velocity components in individual polarity regions, which are sub-regions 
of the active region, have significant variation with time and depth.
In the case of AR 11158, the leading and trailing polarity regions move faster than the
mixed-polarity region. However,  the region-to-region variation is small
when there is no major change in the morphology of these regions. 
Further, both zonal and meridional components have opposite signs for
trailing and leading polarity regions at all depths. These velocity components
 change sign in the mixed-polarity region after the X2.2 class flare on Feb 15,
and in  the trailing-polarity region after the M6.6 flare on Feb 13 that produced  twists
in flow fields of these regions.  We also find sharp dips in the total velocity
in deeper layers around the time of flares.  It is suggested that the  re-organization 
of magnetic fields during flares combined with the sunspot rotation may have reduced 
the magnitude of horizontal flows or that the flow kinetic energy has been converted
into the energy released by flares. After the decline in flare activity and  the sunspot 
rotation,  the flows tend to follow the pattern of the magnetic activity. Finally, the 
velocity components have less variation  near the surface but these variation tend to 
increase with depth demonstrating that the deeper layers are more affected by the topology
of active regions. We plan to identify more active regions with/without different 
characteristics and perform  a statistical study to develop a better picture of the
horizontal flows in the sub-surface layers of complex active regions.

\acknowledgments

{\it SDO} data courtesy of SDO (NASA) and the HMI and AIA consortium. This work was partially supported 
by NASA grant NNH12AT11I and NSF Award 1062054 to the National Solar Observatory. The ring-diagram 
analysis was carried out using the NSO/GONG ring-diagram pipeline. This work was performed 
under the auspices of the SPACEINN Framework of the European Union (EU FP7).

\clearpage

\clearpage


\clearpage


\clearpage


\clearpage


\clearpage


\clearpage

\clearpage
   

\clearpage
   
\clearpage

\begin{table}
\begin{center}

\caption{Evolution of AR 11158 and associated flares on the Earth-facing side of the solar disk\tablenotemark{a}.
\label{table1} }
\begin{tabular}{ccccc}
\tableline\tableline
Date  &  {AR Type\tablenotemark{b}} &\multicolumn{2}{c}{ Sunspot}  &   Number of \\ 
      &                             &    {Area\tablenotemark{c}}           &Count & Flares \\
\hline
2011-02-11 &$\beta$ & 40 & 5 & - \\
2011-02-12 &$\beta$ & 40 & 12 & - \\
2011-02-13 &$\beta\gamma$ & 120 & 31 & 1 M(6.6), 2 C \\
2011-02-14 &$\beta\gamma$ & 450 & 360 & 1 M (2.2),  10 C \\
2011-02-15 &$\beta\gamma\delta$ & 600 & 36 &1 X(2.2), 4 C  \\
2011-02-16 &$\beta\gamma\delta$ & 620 & 19 & 2 M(1.0, 1.6), 9 C \\
2011-02-17 &$\beta\gamma\delta$ & 290 & 17 & 11 C \\
2011-02-18 &$\beta\gamma\delta$ & 310 & 25 & 5 C, 2 M(6.6, 1.4)\\
2011-02-19 &$\beta\gamma$ & 280 & 10 & 7 C\\
2011-02-20 &$\beta\gamma$ & 200 & 10 & -\\
\tableline
\end{tabular}
\tablenotetext{a} {Source: NOAA/SWPC}
\tablenotetext{b}{Magnetic configuration of the spots in the Mt. Wilson system}
\tablenotetext{c}{Area of spots in millionths of the visible hemisphere}

\end{center}
\end{table}
\clearpage

\begin{table}
\begin{center}
\caption{Disk position of various regions analyzed in this study. The analysis is carried out 
using 1920 45-s Dopplergrams for each day. The locations are listed for the reference
image in each time series. \label{table2}  }
\label{tlab}
\begin{tabular}{ccccccrrc}
\tableline\tableline
Region & Region  & Day   &   \multicolumn{2}{c} {Date} &   \multicolumn{2}{c} {Location}& CR \\
Size   & Number & number&    {Quiet\tablenotemark{a}}   & {Active\tablenotemark{b}} &  Long   & Lat   & Long \\
\hline
12$^{\mathrm{o}}$ &--&1& 15-Jan-11 & 11-Feb-11 &  -34.00$^{\mathrm{o}}$ & -20.0$^{\mathrm{o}}$ & 35.07$^{\mathrm{o}}$\\
&&2& 16-Jan-11 & 12-Feb-11 &  -20.80$^{\mathrm{o}}$ & -20.0$^{\mathrm{o}}$ & 35.07$^{\mathrm{o}}$\\
&&3& 17-Jan-11 & 13-Feb-11 &   -7.66$^{\mathrm{o}}$ & -20.0$^{\mathrm{o}}$ & 35.07$^{\mathrm{o}}$\\
&&4& 18-Jan-11 & 14-Feb-11 &  5.50$^{\mathrm{o}}$ & -20.0$^{\mathrm{o}}$ & 35.07$^{\mathrm{o}}$\\
&&5& 19-Jan-11 & 15-Feb-11 &  18.60$^{\mathrm{o}}$ & -20.0$^{\mathrm{o}}$ & 35.07$^{\mathrm{o}}$\\
&&6& 20-Jan-11 & 16-Feb-11 &  31.80$^{\mathrm{o}}$ & -20.0$^{\mathrm{o}}$ & 35.07$^{\mathrm{o}}$\\
\hline
7.5$^{\mathrm{o}}$& R1&1& 15-Jan-11 & 11-Feb-11 &  -37.60$^{\mathrm{o}}$ & -20.5$^{\mathrm{o}}$ & 31.47$^{\mathrm{o}}$\\
&&2& 16-Jan-11 & 12-Feb-11 &  -24.40$^{\mathrm{o}}$ & -20.5$^{\mathrm{o}}$ & 31.47$^{\mathrm{o}}$\\
&&3& 17-Jan-11 & 13-Feb-11 &   -11.20$^{\mathrm{o}}$ & -20.5$^{\mathrm{o}}$ & 31.47$^{\mathrm{o}}$\\
& &4& 18-Jan-11 & 14-Feb-11 &  1.90$^{\mathrm{o}}$ & -20.5$^{\mathrm{o}}$ & 31.47$^{\mathrm{o}}$\\
 &&5& 19-Jan-11 & 15-Feb-11 &  15.00$^{\mathrm{o}}$ & -20.5$^{\mathrm{o}}$ &31.47$^{\mathrm{o}}$\\
&&6& 20-Jan-11 & 16-Feb-11 &   28.20$^{\mathrm{o}}$ & -20.5$^{\mathrm{o}}$ & 31.47$^{\mathrm{o}}$\\
\hline
7.5$^{\mathrm{o}}$& R2&1& 15-Jan-11 & 11-Feb-11 &  -34.00$^{\mathrm{o}}$ & -20.0$^{\mathrm{o}}$ & 35.07$^{\mathrm{o}}$\\
&&2& 16-Jan-11 & 12-Feb-11 &  -20.80$^{\mathrm{o}}$ & -20.0$^{\mathrm{o}}$ & 35.07$^{\mathrm{o}}$\\
&&3& 17-Jan-11 & 13-Feb-11 &   -7.66$^{\mathrm{o}}$ & -20.0$^{\mathrm{o}}$ & 35.07$^{\mathrm{o}}$\\
& &4& 18-Jan-11 & 14-Feb-11 &  5.50$^{\mathrm{o}}$ & -20.0$^{\mathrm{o}}$ & 35.07$^{\mathrm{o}}$\\
 &&5& 19-Jan-11 & 15-Feb-11 &  18.60$^{\mathrm{o}}$ & -20.0$^{\mathrm{o}}$ &35.07$^{\mathrm{o}}$\\
&&6& 20-Jan-11 & 16-Feb-11 &   31.80$^{\mathrm{o}}$ & -20.0$^{\mathrm{o}}$ & 35.07$^{\mathrm{o}}$\\
\hline
7.5$^{\mathrm{o}}$& R3&1& 15-Jan-11 & 11-Feb-11 &  -29.70$^{\mathrm{o}}$ & -19.5$^{\mathrm{o}}$ & 39.37$^{\mathrm{o}}$\\
&&2& 16-Jan-11 & 12-Feb-11 &  -16.50$^{\mathrm{o}}$ & -19.5$^{\mathrm{o}}$ & 39.37$^{\mathrm{o}}$\\
&&3& 17-Jan-11 & 13-Feb-11 &   -3.36$^{\mathrm{o}}$ & -19.5$^{\mathrm{o}}$ & 39.37$^{\mathrm{o}}$\\
& &4& 18-Jan-11 & 14-Feb-11 &  9.80$^{\mathrm{o}}$ & -19.5$^{\mathrm{o}}$ & 39.37$^{\mathrm{o}}$\\
 &&5& 19-Jan-11 & 15-Feb-11 &  22.90$^{\mathrm{o}}$ & -19.5$^{\mathrm{o}}$ &39.37$^{\mathrm{o}}$\\
&&6& 20-Jan-11 & 16-Feb-11 &   36.10$^{\mathrm{o}}$ & -19.5$^{\mathrm{o}}$ & 39.37$^{\mathrm{o}}$\\
\tableline
\end{tabular}
\tablenotetext{a}{03:44:15 UT}
\tablenotetext{b}{11:59:15 UT}
\end{center}

\end{table}

\clearpage

\begin{table}
\begin{center}

\caption{ Mean magnetic activity index (MAI) and standard deviation for different quiet and active regions.  \label{table3} }
\begin{tabular}{ccccccrr}
\tableline\tableline
Region & Region  &    \multicolumn{2}{c} {Quiet Region} &   \multicolumn{2}{c} {Active Region} \\
Size   & Number &     Mean  (G)  & STDDEV (\%)      &     Mean  (G)  & STDDEV  (\%)   \\          
\tableline
12$^{\mathrm{o}}$ &--& 1.4 & 14 &82.1& 57 \\
7.5$^{\mathrm{o}}$& R1& 1.4 & 24 & 118.4& 58  \\
7.5$^{\mathrm{o}}$& R2& 1.4 & 6 & 120.6 &53  \\
7.5$^{\mathrm{o}}$& R3& 1.7 & 26 & 81.2 & 60  \\

\tableline
\end{tabular}
\end{center}
\end{table}

\bibliography{Jain}

\end{document}